\documentclass[12pt]{article}

\newcommand{\Tr}{\mathop{\mathrm{Tr}}\nolimits}

\begin{document}

\title{A new method for calculation of traces of Dirac $\gamma$-matrices
in Minkowski space}

\author {Alexander~L.~Bondarev
\and \it National Scientific and Educational Center of Particle and
\and \it High Energy Physics of the Belarusian State University \and
\rm \rm e-mail: bondarev@hep.by}

\date{}
\maketitle


\begin{abstract}
This paper presents some relations for orthonormal bases in
the Minkowski space and isotropic tetrads constructed from the
vectors of these bases.

As an example of an application of the obtained formulae,
in particular recursion relations, a new method is
proposed to calculate traces of Dirac $\gamma$-matrices in the
Minkowski space. Compared to the classical algorithms, the new method
results in more compact expressions for the traces. Specifically,
it may be easily implemented as a simple yet efficient computer
algorithm.
\end{abstract}


\section{Introduction}
%

The problems arising when the traces of a large
number of Dirac $\gamma$-matrices should be calculated are well known.  For ten
matrices and more the expressions for the traces become extremely
cumbersome.

For example, the classical algorithm based on the application of
the algebra of $\gamma$-matrices
\begin{equation}
\gamma^{\mu} \gamma^{\nu} + \gamma^{\nu} \gamma^{\mu} = 2 g^{\mu
\nu}
\label{e1.1}
\end{equation}
for the trace of ten $\gamma$-matrices results in the expression
consisting of 945 terms (e.g. such algorithm is used in
REDUCE~\cite{r1}).

In four-dimensional space-time somewhat shorter results provides
the method using the Chisholm identities \cite{r2} and Kahane
algorithm \cite{r3} that is based on the reduction formula
\begin{equation}
\gamma^{\mu} \gamma^{\nu} \gamma^{\rho} = g^{\mu \nu} \gamma^{\rho}
- g^{\mu \rho} \gamma^{\nu} + g^{\nu \rho} \gamma^{\mu} + i
\varepsilon^{\mu \nu \rho \lambda} \gamma^5 \gamma_{\lambda} \; .
\label{e1.2}
\end{equation}
This method, implemented first in SCHOONSCHIP~\cite{r4} and later
in FORM~\cite{r5}, gives for the trace of ten $\gamma$-matrices
the following expression consisting of 693 terms:
\begin{eqnarray*}
&& \Tr_4 ( \hat{a}_1 \hat{a}_2 \hat{a}_3 \hat{a}_4 \hat{a}_5 \hat{a}_6
\hat{a}_7 \hat{a}_8 \hat{a}_9 \hat{a}_{10} )
 = \Tr ( \hat{a}_1 \hat{a}_2 \hat{a}_3 \hat{a}_4 \hat{a}_5 \hat{a}_6
\hat{a}_7 \hat{a}_8 \hat{a}_9 \hat{a}_{10} )
          \\
&& +
\begin{array}{l}
 G\pmatrix{ a_1 & a_2 & a_3 & a_6 & a_7 \\ a_4 & a_5 & a_8 & a_9 &
a_{10} }
\end{array}
 -
\begin{array}{l}
 G\pmatrix{ a_1 & a_2 & a_3 & a_8 & a_9 \\ a_4 & a_5 & a_6 & a_7 &
a_{10} }
\end{array}
          \\
&& +
\begin{array}{l}
 G\pmatrix{ a_1 & a_2 & a_3 & a_8 & a_{10} \\ a_4 & a_5 & a_6 & a_7 &
a_9 }
\end{array}
 -
\begin{array}{l}
 G\pmatrix{ a_1 & a_2 & a_3 & a_9 & a_{10} \\ a_4 & a_5 & a_6 & a_7 &
a_8 }
\end{array}
 \;\; .
\end{eqnarray*}

The proposed new method results in more compact expressions (e.g. consisting of 32 terms for
the trace of ten $\gamma$-matrices) than
the classical methods. In particular, this method may be
easily implemented as a simple yet efficient computer algorithm.

The method is based on the properties of orthonormal bases in the
Minkowski space and isotropic tetrads (see e.g. \cite{r6}) constructed from the vectors of
these bases.

In this paper we use the Feynman metrics:
\[  \mu = 0,1,2,3, \;\;\; a^{\mu} = ( a_0 , \vec{a} ) ,
\;\;\; a_{\mu} = ( a_0, -\vec{a} ) , \;\;\; (a b) = a_{\mu} b^{\mu}
= a_0 b_0 - \vec{a} \vec{b} \; , \]
and a sign of the Levi-Civita tensor is determined as $
\; {\varepsilon}_{ 0 1 2 3 } = + 1 $;
\[  \;\;  \hat{a} = \gamma_{\mu} a^{\mu} \; , \;\;
\gamma^5 =
 i \gamma^0 \gamma^1 \gamma^2 \gamma^3 \; .\]

Together with the regular Gram determinants, we use the generalized
ones.  For example:
\[
\begin{array}{l}
G\pmatrix{ {}^{\mu} & {}^{\alpha} \\ {}^{\nu} & {}^{\beta} } =
\left| \matrix{  g^{\mu \nu} & g^{\mu \beta} \\
 g^{\alpha \nu} & g^{\alpha \beta} } \right|
\end{array}
 =  g^{\mu \nu} g^{\alpha
 \beta} - g^{\mu \beta} g^{\alpha \nu}
 ,
\; \; g^{\mu \nu} = \left\{
\begin{array}{rl}
            1, &  \; \mu = \nu = 0  \\
           -1, &  \; \mu = \nu = 1,2,3  \\
            0, &  \; \mu \neq \nu
\end{array}
\right.
\]
$
\begin{array}{l}
G\pmatrix{ {}^{\mu} & {}^{\alpha} \\ {}^{\nu} & {}^{\beta} }
a_{\alpha} b_{\beta} = G\pmatrix{ {}^{\mu} & a \\ {}^{\nu} & b } =
\left| \matrix{ g^{\mu \nu} & b^{\mu} \\ a^{\nu} & (a b) } \right| =
g^{\mu \nu} (a b)  -  b^{\mu} a^{\nu}
\end {array}
$.
%


\section{Some relations for isotropic tetrads in the Minkowski space}
%

Let vectors $l_0$, $l_1$, $l_2$, $l_3$ form an orthonormal basis
in the Minkowski space, i.e.
\[ (l_A l_B) =  g_{A B} \; , \; \; (A,B = 0,1,2,3) \]
(index of the basis vector is denoted by the capital letters).

In four-dimensional space both the regular and generalized fifth-order Gram
determinants equal zero, therefore
\begin{eqnarray}
&&
\begin{array}{l}
G\pmatrix{ {}^{\mu} & l_0 & l_1 & l_2 & l_3 \\
           {}^{\nu} & l_0 & l_1 & l_2 & l_3  }
\end{array}
 =
\begin{array}{l}
\left| \matrix{ g^{\mu \nu} & l_0^{\mu} & l_1^{\mu} & l_2^{\mu} &
l_3^{\mu} \\ l_0^{\nu} & +1 & 0 & 0 & 0 \\ l_1^{\nu} & 0 & -1 & 0
& 0  \\ l_2^{\nu} & 0 & 0 & -1 & 0 \\ l_3^{\nu} & 0 & 0 & 0 & -1}
\right|
\end{array}
   \nonumber       \\
&& = - g^{\mu \nu} + l_0^{\mu} l_0^{\nu} - l_1^{\mu} l_1^{\nu} -
l_2^{\mu} l_2^{\nu} - l_3^{\mu} l_3^{\nu} = 0 \; .
\label{e2.1}
\end{eqnarray}

Whence we obtain the expression for the metric tensor:
\begin{equation}
g^{\mu \nu} =  l_0^{\mu} l_0^{\nu} - l_1^{\mu} l_1^{\nu} -
l_2^{\mu} l_2^{\nu} - l_3^{\mu} l_3^{\nu} \; .
\label{e2.2}
\end{equation}

Furthermore, for the orthonormal basis we have
\begin{equation}
\begin{array}{l}
G\pmatrix{l_0 & l_1 & l_2 & l_3 \\
          l_0 & l_1 & l_2 & l_3  }
\end{array}
=
\begin{array}{l}
\left| \matrix{ +1 & 0 & 0 & 0 \\ 0 & -1 & 0 & 0  \\ 0 & 0 & -1 &
0 \\ 0 & 0 & 0 & -1} \right|
\end{array}
= -1 \; .
\label{e2.3}
\end{equation}

Since
\begin{equation}
\begin{array}{l}
G\pmatrix{l_0 & l_1 & l_2 & l_3 \\
          l_0 & l_1 & l_2 & l_3  }
          \end{array}
= - \varepsilon_{\mu \nu \rho \lambda} l_0^{\mu} l_1^{\nu}
l_2^{\rho} l_3^{\lambda} \ \varepsilon_{\alpha \beta \sigma \tau}
l_0^{\alpha} l_1^{\beta} l_2^{\sigma} l_3^{\tau}  \; ,
\label{e2.4}
\end{equation}
we get:
\begin{equation}
\varepsilon_{\mu \nu \rho \lambda} l_0^{\mu} l_1^{\nu} l_2^{\rho}
l_3^{\lambda} = \pm 1  \; .
\label{e2.5}
\end{equation}
The ``$+$'' and ``$-$'' signs correspond to the ``right-hand''
and ``left-hand'' quadruples of the vectors.

For definiteness in the following we consider the
``right-hand'' basis.
\begin{eqnarray}
{\hat l}_0 {\hat l}_1 {\hat l}_2 {\hat l}_3
& = & {1 \over 4} \left[ \Tr
\left( {\hat l}_0 {\hat l}_1 {\hat l}_2 {\hat l}_3 \right) +
\gamma^5 \Tr \left( \gamma^5 {\hat l}_0 {\hat l}_1 {\hat l}_2 {\hat
l}_3 \right) + {1 \over 2} \sigma^{\mu \nu} \Tr \left( \sigma_{\mu
\nu} {\hat l}_0 {\hat l}_1 {\hat l}_2 {\hat l}_3 \right) \right]
      \nonumber    \\
& = & {1 \over 4} \gamma^5 \Tr \left( \gamma^5 {\hat l}_0 {\hat l}_1
{\hat l}_2 {\hat l}_3 \right) = i \gamma^5 \varepsilon_{\mu \nu \rho
\lambda} l_0^{\mu} l_1^{\nu} l_2^{\rho} l_3^{\lambda} = i \gamma^5
\, ,
\label{e2.6}
\end{eqnarray}
which gives us additional relations for the basis vectors, for example:
\begin{equation}
{\hat l}_0 = i \gamma^5 {\hat l}_1 {\hat l}_2 {\hat l}_3  \; , \;\;
{\hat l}_1 = i \gamma^5 {\hat l}_0 {\hat l}_2 {\hat l}_3 \; , \;\;
{\hat l}_2 = - i  \gamma^5 {\hat l}_0 {\hat l}_1 {\hat l}_3 \; ,
\;\;
{\hat l}_3 = i  \gamma^5 {\hat l}_0 {\hat l}_1 {\hat l}_2 \; .
\label{e2.7}
\end{equation}

Using (\ref{e2.7}), we derive the expressions which we need in
the following:
\begin{equation}
{\hat l}_0 {\hat l}_1 = - i \gamma^5  {\hat l}_2 {\hat l}_3  \; ,
\label{e2.8}
\end{equation}
\begin{equation}
{\hat l}_0 {\hat l}_2 =  i \gamma^5  {\hat l}_1 {\hat l}_3 \; ,
\label{e2.9}
\end{equation}
\begin{equation}
{\hat l}_0 {\hat l}_3 = - i \gamma^5  {\hat l}_1 {\hat l}_2 \; ,
\label{e2.10}
\end{equation}
\begin{equation}
{\hat l}_1 {\hat l}_2 = i \gamma^5  {\hat l}_0 {\hat l}_3  \; ,
\label{e2.11}
\end{equation}
\begin{equation}
{\hat l}_1 {\hat l}_3 = - i \gamma^5  {\hat l}_0 {\hat l}_2 \; ,
\label{e2.12}
\end{equation}
\begin{equation}
{\hat l}_2 {\hat l}_3 = i \gamma^5  {\hat l}_0 {\hat l}_1 \; .
\label{e2.13}
\end{equation}

Let us introduce the isotropic tetrad
\begin{equation}
 q_{+} = { l_0 + l_1 \over \sqrt{2} } \; , \;\;
 q_{-} = { l_0 - l_1 \over \sqrt{2} } \; , \;\;
 (q_{+} q_{-}) = 1 \; ;
\label{e2.14}
\end{equation}
\begin{equation}
 e_{+} = { l_2 + i l_3 \over \sqrt{2} } \; , \;\;
 e_{-} = { l_2 - i l_3 \over \sqrt{2} } \; , \;\;
 (e_{+} e_{-}) = -1 \; , \;\;
 (e_{\mp})^{\ast} = e_{\pm} \; ;
\label{e2.15}
\end{equation}
\[
(q_{+} e_{+}) = (q_{+} e_{-}) = (q_{-} e_{+}) = (q_{-} e_{-}) = 0
 \; .
\]

Then (\ref{e2.2}) takes the form
\begin{equation}
g^{\mu \nu} =  q_{+}^{\mu} q_{-}^{\nu} + q_{-}^{\mu} q_{+}^{\nu} -
e_{+}^{\mu} e_{-}^{\nu} - e_{-}^{\mu} e_{+}^{\nu} \; ,
\label{e2.16}
\end{equation}
from which follow the expressions below:
\begin{equation}
(a b) =  (a q_{+}) (b q_{-}) + (a q_{-}) (b q_{+}) - (a e_{+}) (b
e_{-}) - (a e_{-}) (b e_{+}) \; ,
\label{e2.17}
\end{equation}
\begin{equation}
{\gamma}^{\mu} =   q_{+}^{\mu} \hat{q}_{-} + q_{-}^{\mu} \hat{q}_{+}
- e_{+}^{\mu} \hat{e}_{-} - e_{-}^{\mu} \hat{e}_{+} \; ,
\label{e2.18}
\end{equation}
\begin{equation}
\hat{a} =  (a q_{+}) \hat{q}_{-} + (a q_{-}) \hat{q}_{+} - (a e_{+})
\hat{e}_{-} - (a e_{-}) \hat{e}_{+} \; .
\label{e2.19}
\end{equation}

From (\ref{e2.8})~--~(\ref{e2.13}) we have
\begin{eqnarray}
{\hat q}_{+} {\hat e}_{-} &=& {1 \over 2} ( {\hat l}_0 + {\hat l}_1 )
( {\hat l}_2 - i {\hat l}_3 ) =
 {1 \over 2} ( {\hat l}_0 {\hat l}_2 + {\hat l}_1 {\hat l}_2 - i {\hat
 l}_0 {\hat l}_3 - i {\hat l}_1 {\hat l}_3 )
      \nonumber    \\
&=& {1 \over 2} ( i \gamma^5 {\hat l}_1 {\hat l}_3 + i \gamma^5 {\hat
l}_0 {\hat l}_3 - \gamma^5 {\hat l}_1 {\hat l}_2 -  \gamma^5 {\hat
l}_0 {\hat l}_2 )
      \nonumber    \\
 &=& - \gamma^5 {1 \over 2} ( {\hat l}_0 {\hat l}_2 + {\hat l}_1 {\hat
l}_2 - i {\hat l}_0 {\hat l}_3 - i {\hat l}_1 {\hat l}_3 ) = -
\gamma^5 {\hat q}_{+} {\hat e}_{-} \; .
\label{e2.20}
\end{eqnarray}
In a similar manner
\begin{equation}
{\hat q}_{-} {\hat e}_{+} = - \gamma^5 {\hat q}_{-} {\hat e}_{+}
 \, .
\label{e2.21}
\end{equation}
\begin{equation}
{\hat q}_{+} {\hat e}_{+} = \gamma^5 {\hat q}_{+} {\hat e}_{+}
 \, .
\label{e2.22}
\end{equation}
\begin{equation}
{\hat q}_{-} {\hat e}_{-} = \gamma^5 {\hat q}_{-} {\hat e}_{-}
 \, .
\label{e2.23}
\end{equation}

From (\ref{e2.20}) -- (\ref{e2.23}) it follows that
\begin{equation}
(1 + \gamma^5) \hat{q}_{+} \hat{e}_{-} = 0 \; ,
\label{e2.24}
\end{equation}
\begin{equation}
(1 + \gamma^5) \hat{q}_{-} \hat{e}_{+} = 0 \; ,
\label{e2.25}
\end{equation}
\begin{equation}
(1 - \gamma^5) \hat{q}_{+} \hat{e}_{+} = 0 \; ,
\label{e2.26}
\end{equation}
\begin{equation}
(1 - \gamma^5) \hat{q}_{-} \hat{e}_{-} = 0 \; .
\label{e2.27}
\end{equation}

Hence
\begin{equation}
(1 + \gamma^5) \hat{e}_{-} \hat{e}_{+} \hat{q}_{-}
 = - 2 (1 + \gamma^5) \hat{q}_{-} \; ,
\label{e2.28}
\end{equation}
\begin{equation}
(1 + \gamma^5) \hat{e}_{+} \hat{e}_{-} \hat{q}_{+}
 = - 2 (1 + \gamma^5) \hat{q}_{+} \; ,
\label{e2.29}
\end{equation}
\begin{equation}
(1 + \gamma^5) \hat{q}_{-} \hat{q}_{+} \hat{e}_{-}
 = 2 (1 + \gamma^5) \hat{e}_{-} \; ,
\label{e2.30}
\end{equation}
\begin{equation}
(1 + \gamma^5) \hat{q}_{+} \hat{q}_{-} \hat{e}_{+}
 = 2 (1 + \gamma^5) \hat{e}_{+} \; .
\label{e2.31}
\end{equation}
\begin{equation}
(1 - \gamma^5) \hat{e}_{+} \hat{e}_{-} \hat{q}_{-}
 = - 2 (1 - \gamma^5) \hat{q}_{-} \; ,
\label{e2.32}
\end{equation}
\begin{equation}
(1 - \gamma^5) \hat{e}_{-} \hat{e}_{+} \hat{q}_{+}
 = - 2 (1 - \gamma^5) \hat{q}_{+} \; ,
\label{e2.33}
\end{equation}
\begin{equation}
(1 - \gamma^5) \hat{q}_{+} \hat{q}_{-} \hat{e}_{-}
 = 2 (1 - \gamma^5) \hat{e}_{-} \; ,
\label{e2.34}
\end{equation}
\begin{equation}
(1 - \gamma^5) \hat{q}_{-} \hat{q}_{+} \hat{e}_{+}
 = 2 (1 - \gamma^5) \hat{e}_{+} \; .
\label{e2.35}
\end{equation}

For example, for (\ref{e2.28}) we have
\[
(1 + \gamma^5) \hat{e}_{-} \hat{e}_{+} \hat{q}_{-}
 = 2 (e_{-} e_{+}) (1 + \gamma^5) \hat{q}_{-}
 - (1 + \gamma^5) \hat{e}_{+} \hat{e}_{-} \hat{q}_{-}\; ,
\]
where the second term on the right-hand side equals zero due to
(\ref{e2.27}).

Furthermore, from the definition of the isotropic
tetrad (\ref{e2.14}), (\ref{e2.15}) it is clear that
\begin{equation}
\hat{q}_{-} \hat{q}_{+} \hat{q}_{-} = 2 \hat{q}_{-} \; ,
\label{e2.36}
\end{equation}
\begin{equation}
\hat{q}_{+} \hat{q}_{-} \hat{q}_{+} = 2 \hat{q}_{+} \; ,
\label{e2.37}
\end{equation}
\begin{equation}
\hat{e}_{-} \hat{e}_{+} \hat{e}_{-} = - 2 \hat{e}_{-} \; ,
\label{e2.38}
\end{equation}
\begin{equation}
\hat{e}_{+} \hat{e}_{-} \hat{e}_{+} = - 2 \hat{e}_{+} \; .
\label{e2.39}
\end{equation}

It is important, that any three
operators of the form $\hat{q}_{\lambda}$ or $\hat{e}_{\sigma}$ taken
in succession in the presence of $(1 \pm \gamma^5)$ are either reduced to
only one of them or equal zero.


\section{Recursion relations}
%

Now, using (\ref{e2.19}), (\ref{e2.24}) -- (\ref{e2.39}) we
have
\begin{eqnarray}
&& (1 + \gamma^5) \hat{q}_{-} \hat{a} \hat{b}
 = (a q_{-}) (1 + \gamma^5) \hat{q}_{-} \hat{q}_{+} \hat{b}
 - (a e_{+}) (1 + \gamma^5) \hat{q}_{-} \hat{e}_{-} \hat{b}
    \nonumber \\[0.2cm]
&=& (a q_{-}) [ (b q_{+}) (1 + \gamma^5) \hat{q}_{-} \hat{q}_{+}
\hat{q}_{-} - (b e_{+}) (1 + \gamma^5) \hat{q}_{-} \hat{q}_{+}
\hat{e}_{-} ]
    \nonumber  \\
&-& (a e_{+}) [ (b q_{-}) (1 + \gamma^5) \hat{q}_{-} \hat{e}_{-}
\hat{q}_{+} - (b e_{-}) (1 + \gamma^5) \hat{q}_{-} \hat{e}_{-}
\hat{e}_{+} ]
     \nonumber  \\[0.2cm]
&=& (a q_{-}) 2 [ (b q_{+}) (1 + \gamma^5) \hat{q}_{-} - (b e_{+}) (1
+ \gamma^5) \hat{e}_{-} ]
    \nonumber  \\
 &-& (a e_{+}) 2 [ - (b q_{-}) (1 + \gamma^5)
\hat{e}_{-} + (b e_{-}) (1 + \gamma^5) \hat{q}_{-} ]
    \nonumber  \\[0.2cm]
&=& 2 [ (a q_{-}) (b q_{+}) - (a e_{+}) (b e_{-}) ] (1 + \gamma^5)
\hat{q}_{-}
    \nonumber  \\
 &+& 2 [ (a e_{+}) (b q_{-}) - (a q_{-}) (b e_{+}) ] (1 + \gamma^5)
\hat{e}_{-} \; .
\label{e3.1}
\end{eqnarray}
In the same way
\begin{eqnarray}
&& (1 + \gamma^5) \hat{e}_{-} \hat{a} \hat{b}
 = (a q_{+}) (1 + \gamma^5) \hat{e}_{-} \hat{q}_{-} \hat{b}
 - (a e_{-}) (1 + \gamma^5) \hat{e}_{-} \hat{e}_{+} \hat{b}
     \nonumber  \\[0.2cm]
&=& (a q_{+}) [ (b q_{-}) (1 + \gamma^5) \hat{e}_{-} \hat{q}_{-}
\hat{q}_{+} - (b e_{-}) (1 + \gamma^5) \hat{e}_{-} \hat{q}_{-}
\hat{e}_{+} ]
     \nonumber  \\
&-& (a e_{-}) [ (b q_{+}) (1 + \gamma^5) \hat{e}_{-} \hat{e}_{+}
\hat{q}_{-} - (b e_{+}) (1 + \gamma^5) \hat{e}_{-} \hat{e}_{+}
\hat{e}_{-} ]
     \nonumber  \\[0.2cm]
&=& (a q_{+}) 2 [ (b q_{-}) (1 + \gamma^5) \hat{e}_{-} - (b e_{-}) (1
+ \gamma^5) \hat{q}_{-} ]
     \nonumber  \\
&-& (a e_{-}) 2 [ - (b q_{+}) (1 + \gamma^5) \hat{q}_{-} + (b e_{+})
(1 + \gamma^5) \hat{e}_{-} ]
     \nonumber  \\[0.2cm]
&=& 2 [ (a q_{+}) (b q_{-}) - (a e_{-}) (b e_{+}) ] (1 + \gamma^5)
\hat{e}_{-}
     \nonumber  \\
&+& 2 [ (a e_{-}) (b q_{+}) - (a q_{+}) (b e_{-}) ] (1 + \gamma^5)
\hat{q}_{-} \; .
\label{e3.2}
\end{eqnarray}

Defining the functions
\begin{equation}
F_1 (a, b) = 2 [ (a q_{-}) (b q_{+}) - (a e_{+}) (b e_{-}) ] \; ,
\label{e3.3}
\end{equation}
\begin{equation}
F_2 (a, b) = 2 [ (a e_{+}) (b q_{-}) - (a q_{-}) (b e_{+}) ]
 = 2
\begin {array}{l}
 G\pmatrix{ a & b \\ e_{+} & q_{-} }
\end {array}
\; ,
\label{e3.4}
\end{equation}
\begin{equation}
F_3 (a, b) = 2 [ (a q_{+}) (b q_{-}) - (a e_{-}) (b e_{+}) ] \; ,
\label{e3.5}
\end{equation}
\begin{equation}
F_4 (a, b) = 2 [ (a e_{-}) (b q_{+}) - (a q_{+}) (b e_{-}) ]
= 2
\begin {array}{l}
 G\pmatrix{ a & b \\ e_{-} & q_{+} }
\end {array}
\; ,
\label{e3.6}
\end{equation}
we finally get
\begin{equation}
(1 + \gamma^5) \hat{q}_{-} \hat{a} \hat{b}
 = F_1 (a, b) (1 + \gamma^5) \hat{q}_{-}
 + F_2 (a, b) (1 + \gamma^5) \hat{e}_{-} \; ,
\label{e3.7}
\end{equation}
\begin{equation}
(1 + \gamma^5) \hat{e}_{-} \hat{a} \hat{b}
 = F_3 (a, b) (1 + \gamma^5) \hat{e}_{-}
 + F_4 (a, b) (1 + \gamma^5) \hat{q}_{-} \; .
\label{e3.8}
\end{equation}

Making similar calculations, we obtain
\begin{equation}
(1 + \gamma^5) \hat{q}_{+} \hat{a} \hat{b}
 = F_3 (a, b) (1 + \gamma^5) \hat{q}_{+}
 + F_4 (a, b) (1 + \gamma^5) \hat{e}_{+} \; ,
\label{e3.9}
\end{equation}
\begin{equation}
(1 + \gamma^5) \hat{e}_{+} \hat{a} \hat{b}
 = F_1 (a, b) (1 + \gamma^5) \hat{e}_{+}
 + F_2 (a, b) (1 + \gamma^5) \hat{q}_{+} \; .
\label{e3.10}
\end{equation}

In a similar manner, after defining the functions
\begin{equation}
F_5 (a, b) = 2 [ (a q_{-}) (b q_{+}) - (a e_{-}) (b e_{+}) ] \; ,
\label{e3.11}
\end{equation}
\begin{equation}
F_6 (a, b) = 2 [ (a e_{-}) (b q_{-}) - (a q_{-}) (b e_{-}) ]
= 2
\begin {array}{l}
 G\pmatrix{ a & b \\ e_{-} & q_{-} }
\end {array}
\; ,
\label{e3.12}
\end{equation}
\begin{equation}
F_7 (a, b) = 2 [ (a q_{+}) (b q_{-}) - (a e_{+}) (b e_{-}) ] \; ,
\label{e3.13}
\end{equation}
\begin{equation}
F_8 (a, b) = 2 [ (a e_{+}) (b q_{+}) - (a q_{+}) (b e_{+}) ]
= 2
\begin {array}{l}
 G\pmatrix{ a & b \\ e_{+} & q_{+} }
\end {array}
\label{e3.14}
\end{equation}
we obtain
\begin{equation}
(1 - \gamma^5) \hat{q}_{-} \hat{a} \hat{b}
 = F_5 (a, b) (1 - \gamma^5) \hat{q}_{-}
 + F_6 (a, b) (1 - \gamma^5) \hat{e}_{+} \; ,
\label{e3.15}
\end{equation}
\begin{equation}
(1 - \gamma^5) \hat{e}_{+} \hat{a} \hat{b}
 = F_7 (a, b) (1 - \gamma^5) \hat{e}_{+}
 + F_8 (a, b) (1 - \gamma^5) \hat{q}_{-} \; ,
\label{e3.16}
\end{equation}
\begin{equation}
(1 - \gamma^5) \hat{q}_{+} \hat{a} \hat{b}
 = F_7 (a, b) (1 - \gamma^5) \hat{q}_{+}
 + F_8 (a, b) (1 - \gamma^5) \hat{e}_{-} \; ,
\label{e3.17}
\end{equation}
\begin{equation}
(1 - \gamma^5) \hat{e}_{-} \hat{a} \hat{b}
 = F_5 (a, b) (1 - \gamma^5) \hat{e}_{-}
 + F_6 (a, b) (1 - \gamma^5) \hat{q}_{+} \; .
\label{e3.18}
\end{equation}

Note that the functions defined in (\ref{e3.3}) --
(\ref{e3.6}), (\ref{e3.11}) -- (\ref{e3.14}) are not
independent:
\[
F_5 (a, b) = F_1^{\ast} (a, b) \; ,
\]
\[
F_6 (a, b) = F_2^{\ast} (a, b) \; ,
\]
\[
F_7 (a, b) = F_3^{\ast} (a, b) \; ,
\]
\[
F_8 (a, b) = F_4^{\ast} (a, b) \; .
\]
Furthermore, the following relations hold:
\[
F_3 (b, a) = F_1 (a, b)   \; ,
\]
\[
F_1 (a, b) + F_3 (a, b) = 2 (a b)    \; ,
\]
\[
F_1 (a, b) F_3 (a, b) - F_2 (a, b) F_4 (a, b) = a^2 b^2   
\]
(see also {\bf Appendix B}).


\section{A new method for calculation of traces}
%
We now use the formulae (\ref{e3.3}) -- (\ref{e3.18}) obtained in
the previous section to calculate the traces of Dirac
$\gamma$-matrices as follows:
\begin{eqnarray}
 \Tr [(1 + \gamma^5) \hat{q}_{-} \hat{a}_1 \hat{a}_2 \hat{a}_3 \cdots
\hat{a}_{2n+1} ]
 &=& F_1 (a_1, a_2) \Tr [(1 + \gamma^5) \hat{q}_{-} \hat{a}_3 \cdots
\hat{a}_{2n+1} ]
   \nonumber  \\
 &+& F_2 (a_1, a_2) \Tr [(1 + \gamma^5) \hat{e}_{-} \hat{a}_3 \cdots
\hat{a}_{2n+1} ] \; ,
\label{e4.1}
\end{eqnarray}
\begin{eqnarray}
\Tr [(1 + \gamma^5) \hat{e}_{-} \hat{a}_1 \hat{a}_2 \hat{a}_3 \cdots
\hat{a}_{2n+1} ]
 &=& F_3 (a_1, a_2) \Tr [(1 + \gamma^5) \hat{e}_{-} \hat{a}_3 \cdots
\hat{a}_{2n+1} ]
   \nonumber  \\
 &+& F_4 (a_1, a_2) \Tr [(1 + \gamma^5) \hat{q}_{-} \hat{a}_3 \cdots
\hat{a}_{2n+1} ] \; ,
\label{e4.2}
\end{eqnarray}
\begin{eqnarray}
 \Tr [(1 + \gamma^5) \hat{q}_{+} \hat{a}_1 \hat{a}_2 \hat{a}_3 \cdots
\hat{a}_{2n+1} ]
 &=& F_3 (a_1, a_2) \Tr [(1 + \gamma^5) \hat{q}_{+} \hat{a}_3 \cdots
\hat{a}_{2n+1} ]
   \nonumber  \\
 &+& F_4 (a_1, a_2) \Tr [(1 + \gamma^5) \hat{e}_{+} \hat{a}_3 \cdots
\hat{a}_{2n+1} ] \; ,
\label{e4.3}
\end{eqnarray}
\begin{eqnarray}
\Tr [(1 + \gamma^5) \hat{e}_{+} \hat{a}_1 \hat{a}_2 \hat{a}_3 \cdots
\hat{a}_{2n+1} ]
 &=& F_1 (a_1, a_2) \Tr [(1 + \gamma^5) \hat{e}_{+} \hat{a}_3 \cdots
\hat{a}_{2n+1} ]
   \nonumber  \\
 &+& F_2 (a_1, a_2) \Tr [(1 + \gamma^5) \hat{q}_{+} \hat{a}_3 \cdots
\hat{a}_{2n+1} ] \; ,
\label{e4.4}
\end{eqnarray}
and so forth.

As an example, let us calculate
\[
\Tr [(1 - \gamma^5)  \hat{a}_1 \hat{a}_2 \hat{a}_3 \hat{a}_4
\hat{a}_5 \hat{a}_6 ] \; .
\]

We take decomposition (\ref{e2.19}) written in the form
\begin{eqnarray}
(1 + \gamma^5) \hat{a}_6 &=&
   (1 + \gamma^5) \hat{q}_{-} (a_6 q_{+})
 + (1 + \gamma^5) \hat{q}_{+} (a_6 q_{-})
    \nonumber \\
 &-& (1 + \gamma^5) \hat{e}_{-} (a_6 e_{+})
 - (1 + \gamma^5) \hat{e}_{+} (a_6 e_{-}) \; .
\label{e4.5}
\end{eqnarray}

Using (\ref{e4.1}) and (\ref{e4.2}), we get
\begin{eqnarray*}
&& \Tr [(1 + \gamma^5) \hat{q}_{-} \hat{a}_1 \hat{a}_2 \hat{a}_3
\hat{a}_4 \hat{a}_5 ]
   \nonumber \\[0.1cm]
&=& F_1(a_1, a_2) \Tr [(1 + \gamma^5) \hat{q}_{-} \hat{a}_3 \hat{a}_4
\hat{a}_5 ]
 + F_2(a_1, a_2) \Tr [(1 + \gamma^5) \hat{e}_{-} \hat{a}_3
\hat{a}_4 \hat{a}_5 ]
   \nonumber \\[0.1cm]
&=& F_1(a_1, a_2) \left\{  F_1(a_3, a_4) \Tr [(1 + \gamma^5)
\hat{q}_{-} \hat{a}_5 ]
 + F_2(a_3, a_4) \Tr [(1 + \gamma^5) \hat{e}_{-} \hat{a}_5 ] \right\}
   \nonumber \\
&+& F_2(a_1, a_2) \left\{  F_3(a_3, a_4) \Tr [(1 + \gamma^5)
\hat{e}_{-} \hat{a}_5 ]
 + F_4(a_3, a_4) \Tr [(1 + \gamma^5) \hat{q}_{-} \hat{a}_5 ] \right\}
   \nonumber \\[0.1cm]
&=& [ F_1(a_1, a_2) F_1(a_3, a_4) + F_2(a_1, a_2) F_4(a_3, a_4) ]
\cdot 4 (a_5 q_{-})
   \nonumber \\
&+& [ F_1(a_1, a_2) F_2(a_3, a_4) + F_2(a_1, a_2) F_3(a_3, a_4) ]
\cdot 4 (a_5 e_{-})
\end{eqnarray*}
i.e.
\begin{eqnarray}
&& \Tr [(1 + \gamma^5) \hat{q}_{-} \hat{a}_1 \hat{a}_2 \hat{a}_3
\hat{a}_4 \hat{a}_5 ] (a_6 q_{+})
   \nonumber \\[0.1cm]
&=& 4 [ F_1(a_1, a_2) F_1(a_3, a_4) + F_2(a_1, a_2) F_4(a_3, a_4) ]
\cdot (a_5 q_{-}) (a_6 q_{+})
   \nonumber \\
&+& 4 [ F_1(a_1, a_2) F_2(a_3, a_4) + F_2(a_1, a_2) F_3(a_3, a_4) ]
\cdot (a_5 e_{-}) (a_6 q_{+}) \; .
\label{e4.6a}
\end{eqnarray}

Making similar calculations for the remaining three items in
decomposition (\ref{e4.5}), we obtain
\begin{eqnarray}
&& \Tr [(1 + \gamma^5) \hat{q}_{+} \hat{a}_1 \hat{a}_2 \hat{a}_3
\hat{a}_4 \hat{a}_5 ] (a_6 q_{-})
   \nonumber \\[0.1cm]
&=& 4 [ F_3(a_1, a_2) F_3(a_3, a_4) + F_4(a_1, a_2) F_2(a_3, a_4) ]
\cdot (a_5 q_{+}) (a_6 q_{-})
   \nonumber \\
&+& 4 [ F_3(a_1, a_2) F_4(a_3, a_4) + F_4(a_1, a_2) F_1(a_3, a_4) ]
\cdot (a_5 e_{+}) (a_6 q_{-}) \; ,
\label{e4.6b}
\end{eqnarray}
\begin{eqnarray}
&-& \Tr [(1 + \gamma^5) \hat{e}_{-} \hat{a}_1 \hat{a}_2 \hat{a}_3
\hat{a}_4 \hat{a}_5 ] (a_6 e_{+})
   \nonumber \\[0.1cm]
&=& - 4 [ F_3(a_1, a_2) F_3(a_3, a_4) + F_4(a_1, a_2) F_2(a_3, a_4) ]
\cdot (a_5 e_{-}) (a_6 e_{+})
   \nonumber \\
&-& 4 [ F_3(a_1, a_2) F_4(a_3, a_4) + F_4(a_1, a_2) F_1(a_3, a_4) ]
\cdot (a_5 q_{-}) (a_6 e_{+}) \; ,
\label{e4.6c}
\end{eqnarray}
\begin{eqnarray}
&-& \Tr [(1 + \gamma^5) \hat{e}_{+} \hat{a}_1 \hat{a}_2 \hat{a}_3
\hat{a}_4 \hat{a}_5 ] (a_6 e_{-})
   \nonumber \\[0.1cm]
&=& - 4 [ F_1(a_1, a_2) F_1(a_3, a_4) + F_2(a_1, a_2) F_4(a_3, a_4) ]
\cdot (a_5 e_{+}) (a_6 e_{-})
   \nonumber \\
&-& 4 [ F_1(a_1, a_2) F_2(a_3, a_4) + F_2(a_1, a_2) F_3(a_3, a_4) ]
\cdot (a_5 q_{+}) (a_6 e_{-}) \; ,
\label{e4.6d}
\end{eqnarray}

and finally
\begin{eqnarray}
&& {1 \over 2} \Tr [(1 - \gamma^5) \hat{a}_1 \hat{a}_2 \hat{a}_3
\hat{a}_4 \hat{a}_5 \hat{a}_6 ]
   \nonumber  \\[0.1cm]
&=& [ F_1(a_1, a_2) F_1(a_3, a_4) + F_2(a_1, a_2) F_4(a_3, a_4) ]
F_1(a_5, a_6)
   \nonumber  \\
&+& [ F_1(a_1, a_2) F_2(a_3, a_4) + F_2(a_1, a_2) F_3(a_3, a_4) ]
F_4(a_5, a_6)
   \nonumber  \\
&+& [ F_3(a_1, a_2) F_3(a_3, a_4) + F_4(a_1, a_2) F_2(a_3, a_4) ]
F_3(a_5, a_6)
   \nonumber  \\
&+& [ F_3(a_1, a_2) F_4(a_3, a_4) + F_4(a_1, a_2) F_1(a_3, a_4) ]
F_2(a_5, a_6) \; ,
\label{e4.7}
\end{eqnarray}
that is we derive an expression consisting of eight terms only.

An expression for
\[
\Tr [(1 + \gamma^5) \hat{a}_1 \hat{a}_2 \hat{a}_3 \hat{a}_4
\hat{a}_5 \hat{a}_6 ]
\]
may be obtained from (\ref{e4.7}) by replacing all the functions with
the complex-conjugate ones.

Having calculated
\[
\Tr [(1 - \gamma^5) \hat{a}_1 \cdots \hat{a}_{2n-1} \hat{a}_{2n} ]
\; ,
\]
we may obtain an expression for
\[
\Tr [(1 - \gamma^5) \hat{a}_1 \cdots \hat{a}_{2n-1} \hat{a}_{2n}
\hat{a}_{2n+1} \hat{a}_{2n+2} ]
\]
from the previous one by the following replacement:
\begin{equation}
F_1 (a_{2n-1}, a_{2n})
 \rightarrow F_1 (a_{2n-1}, a_{2n}) F_1 (a_{2n+1}, a_{2n+2})
 + F_2 (a_{2n-1}, a_{2n}) F_4 (a_{2n+1}, a_{2n+2}) \; ,
\label{e4.8}
\end{equation}
\begin{equation}
F_3 (a_{2n-1}, a_{2n})
 \rightarrow F_3 (a_{2n-1}, a_{2n}) F_3 (a_{2n+1}, a_{2n+2})
 + F_4 (a_{2n-1}, a_{2n}) F_2 (a_{2n+1}, a_{2n+2}) \; ,
\label{e4.9}
\end{equation}
\begin{equation}
F_2 (a_{2n-1}, a_{2n})
 \rightarrow F_1 (a_{2n-1}, a_{2n}) F_2 (a_{2n+1}, a_{2n+2})
 + F_2 (a_{2n-1}, a_{2n}) F_3 (a_{2n+1}, a_{2n+2}) \; ,
\label{e4.10}
\end{equation}
\begin{equation}
F_4 (a_{2n-1}, a_{2n})
 \rightarrow F_3 (a_{2n-1}, a_{2n}) F_4 (a_{2n+1}, a_{2n+2})
 + F_4 (a_{2n-1}, a_{2n}) F_1 (a_{2n+1}, a_{2n+2})
\; .
\label{e4.11}
\end{equation}

Actually,
\begin{eqnarray*}
&& F_1 (a_{2n-1}, a_{2n}) = 2 [ (a_{2n-1} q_{-}) (a_{2n} q_{+}) -
(a_{2n-1} e_{+}) (a_{2n} e_{-}) ]
          \\[0.1cm]
&=& {1 \over 2} \Big\{ \Tr [(1 + \gamma^5) \hat{q}_{-} \hat{a}_{2n-1}
] (a_{2n} q_{+}) - \Tr [(1 + \gamma^5) \hat{e}_{+} \hat{a}_{2n-1} ]
(a_{2n} e_{-}) \Big\}
          \\[0.1cm]
&\rightarrow& {1 \over 2} \Big\{ \Tr [(1 + \gamma^5) \hat{q}_{-}
\hat{a}_{2n-1} \hat{a}_{2n} \hat{a}_{2n+1} ] (a_{2n+2} q_{+})
 - \Tr [(1 + \gamma^5) \hat{e}_{+} \hat{a}_{2n-1} \hat{a}_{2n}
\hat{a}_{2n+1} ] (a_{2n+2} e_{-}) \Big\}
          \\[0.1cm]
&=& 2 \Big\{ [ F_1 (a_{2n-1}, a_{2n}) (a_{2n+1} q_{-}) + F_2
(a_{2n-1}, a_{2n}) (a_{2n+1} e_{-}) ] (a_{2n+2} q_{+})
          \\
&-& [ F_1 (a_{2n-1}, a_{2n}) (a_{2n+1} e_{+}) + F_2 (a_{2n-1}, a_{2n})
(a_{2n+1} q_{+}) ] (a_{2n+2} e_{-}) \Big\}
          \\[0.1cm]
&=&  F_1 (a_{2n-1}, a_{2n}) F_1 (a_{2n+1}, a_{2n+2})
 + F_2 (a_{2n-1}, a_{2n}) F_4 (a_{2n+1}, a_{2n+2}) \; ,
\end{eqnarray*}
and the procedure for the remaining three functions is quite similar.

Note that
\begin{equation}
{1 \over 2} \Tr [(1 - \gamma^5) \hat{a}_1 \hat{a}_2 ] = 2 (a_1 a_2)
= F_1(a_1, a_2) + F_3(a_1, a_2)
\; .
\label{e4.12}
\end{equation}
Thus, formulae (\ref{e4.12}), (\ref{e4.8})~--~(\ref{e4.11}) provide a
very simple method of trace calculation
(see {\bf Appendix A}).

In numerical calculations, after fixing a reference system and
choosing an orthonormal basis, for example, of the form
\[
 l^{\mu}_0 = (1,\ 0, 0, 0) \; , \;\;
 l^{\mu}_1 = (0,\ 1, 0, 0) \; , \;\;
 l^{\mu}_2 = (0,\ 0, 1, 0) \; , \;\;
 l^{\mu}_3 = (0,\ 0, 0, 1) \; ,
\]
we get
\[
q^{\mu}_{\pm} = {1 \over \sqrt{2} } (1,\ \pm 1 , 0, 0) \;,
\hspace{1cm} (a q_{\pm} ) = {1 \ \over \sqrt{2} } ( a_0 \mp a_x) \;
,
\]
\[
e^{\mu}_{\pm} = {1 \over \sqrt{2} } (0,\ 0 , 1, \pm i) \;,
\hspace{1cm} (a e_{\pm} ) = - {1 \ \over \sqrt{2} } ( a_y \pm i a_z)
 \; ,
\]
and
\begin{equation}
F_1 (a, b) = (a b) - (a_0 b_x - a_x b_0) + i (a_y b_z - a_z b_y) \;
,
\label{e4.13}
\end{equation}
\begin{equation}
F_3 (a, b) = (a b) + (a_0 b_x - a_x b_0) - i (a_y b_z - a_z b_y) \;
,
\label{e4.14}
\end{equation}
\begin{equation}
F_2 (a, b) = (a_0 b_y - a_y b_0) + i (a_x b_z - a_z b_x)
 + (a_x b_y - a_y b_x) + i (a_0 b_z - a_z b_0) \; ,
\label{e4.15}
\end{equation}
\begin{equation}
F_4 (a, b) = (a_0 b_y - a_y b_0) + i (a_x b_z - a_z b_x)
 - (a_x b_y - a_y b_x) - i (a_0 b_z - a_z b_0) \; ,
\label{e4.16}
\end{equation}
that is we derive a number of the expressions well-suited for the implementation
of a simple yet efficient computer algorithm.

Comparison of different methods for trace calculation is given in Tab.~\ref{t.1}.
%
\begin{table}[h!t]
\caption{Number of terms in the expression for traces} \label{t.1}
\begin{tabular}{| p{4cm} | l || c | c | c | c | c | c | }
\hline \hline
\multicolumn{2}{| c ||}
{Number of $\gamma$-matrices,  $2n$ } & 2 & 4 & 6 & 8 & 10 & 12 \\
\hline \hline
Number of terms
&$\Tr ( \hat{a}_1 \hat{a}_2 \cdots \hat{a}_{2n} )$
& 1 & 3 & 15 & 105 & 945 & 10395
\\ \cline{2-8}
for algorithm (\ref{e1.1})
& $\Tr ( \gamma^5\hat{a}_1 \hat{a}_2 \cdots \hat{a}_{2n} )$
& 0 & 1 & 15 & 210 & 3150 & 51975 \\
\hline \hline
Number of terms
&$\Tr_4 ( \hat{a}_1 \hat{a}_2 \cdots \hat{a}_{2n} )$
& 1 & 3 & 15 & 105 & 693 & 4383
\\ \cline{2-8}
for algorithm (\ref{e1.2})
& $\Tr_4 ( \gamma^5\hat{a}_1 \hat{a}_2 \cdots \hat{a}_{2n} )$
& 0 & 1 & 6 & 33 & 180 & 1029 \\
\hline \hline
Number of terms for new method (\ref{e4.8})--(\ref{e4.12})
& $\Tr [(1 - \gamma^5) \hat{a}_1 \hat{a}_2 \cdots
\hat{a}_{2n} ]$ & 2 & 4 & 8 & 16 & 32 & 64  \\
\hline \hline
\end{tabular}
\end{table}

In case that traces of the form
\[
 \Tr ( {\gamma}_{\rho} \hat{a}_1 \hat{a}_2 \cdots \hat{a}_{2n+1} )
 \Tr ( {\gamma}^{\rho} \hat{b}_1 \hat{b}_2 \cdots \hat{b}_{2m+1} )
\; ,
\]
need to be calculated, one may use the consequence of formula
(\ref{e2.18})
\[
{\gamma}_{\rho} \times {\gamma}^{\rho}
 = \hat{q}_{-} \times \hat{q}_{+}
 + \hat{q}_{+} \times \hat{q}_{-}
 - \hat{e}_{-} \times \hat{e}_{+}
 - \hat{e}_{+} \times \hat{e}_{-}  \; ,
\]
or sum over index $\rho$ using the classical methods, e.g.:
\begin{eqnarray*}
&& \Tr ( {\gamma}_{\rho} \hat{a}_1 \hat{a}_2 \cdots \hat{a}_{2n+1} )
 \Tr ( {\gamma}^{\rho} \hat{b}_1 \hat{b}_2 \cdots \hat{b}_{2m+1} )
          \\
&& = 2 \Tr [ ( \hat{a}_1 \hat{a}_2 \cdots \hat{a}_{2n+1} +
\hat{a}_{2n+1} \cdots \hat{a}_2 \hat{a}_1 )
 \hat{b}_1 \hat{b}_2 \cdots \hat{b}_{2m+1} ] \; .
\end{eqnarray*}

Finally, we would like to note that for traces of four, six and
eight Dirac $\gamma$-matrices we have directly compared the
expressions for the traces obtained both by the new and
classical methods:
\begin{itemize}
\item[-]
expressions for the new method were expanded by means of formulae
(\ref{e4.13})~--~(\ref{e4.16}) for functions $F$;
\item[-]
in expressions for traces obtained by the classical method the
scalar products and contractions of the
Levi-Civita tensors with 4-vectors were written in terms of vector
components.
\end{itemize}
The results obtained in different ways were identical.

Note that some other examples of applications of isotropic
tetrads to the calculations in high energy physics may be found in
\cite{r7}.


\begin{appendix}

\section{Expressions for traces}
%
%
\begin{equation}
{1 \over 2} \Tr [(1 - \gamma^5) \hat{a}_1 \hat{a}_2 ]
= F_1(a_1, a_2) + F_3(a_1, a_2)
\; ;
\label{ea.1}
\end{equation}
\begin{eqnarray}
{1 \over 2} \Tr [(1 - \gamma^5) \hat{a}_1 \hat{a}_2 \hat{a}_3
\hat{a}_4 ]
 &=& F_1(a_1, a_2) F_1(a_3, a_4) + F_2(a_1, a_2) F_4(a_3, a_4)
          \nonumber \\
&+& F_3(a_1, a_2) F_3(a_3, a_4) + F_4(a_1, a_2) F_2(a_3, a_4)
\; ;
\label{ea.2}
\end{eqnarray}
\begin{eqnarray}
& {1 \over 2} & \Tr [(1 - \gamma^5) \hat{a}_1 \hat{a}_2 \hat{a}_3
\hat{a}_4 \hat{a}_5 \hat{a}_6 ]
          \nonumber \\
& = & F_1(a_1, a_2)
 [ F_1(a_3, a_4) F_1(a_5, a_6) + F_2(a_3, a_4) F_4(a_5, a_6) ]
          \nonumber \\
& + & F_2(a_1, a_2)
 [ F_3(a_3, a_4) F_4(a_5, a_6) + F_4(a_3, a_4) F_1(a_5, a_6) ]
          \nonumber \\
& + & F_3(a_1, a_2)
 [ F_3(a_3, a_4) F_3(a_5, a_6) + F_4(a_3, a_4) F_2(a_5, a_6) ]
          \nonumber \\
& + & F_4(a_1, a_2)
 [ F_1(a_3, a_4) F_2(a_5, a_6) + F_2(a_3, a_4) F_3(a_5, a_6) ]
          \nonumber \\[0.1cm]
& = & [ F_1(a_1, a_2) F_1(a_3, a_4) + F_2(a_1, a_2) F_4(a_3, a_4) ]
F_1(a_5, a_6)
          \nonumber \\
& + & [ F_3(a_1, a_2) F_4(a_3, a_4) + F_4(a_1, a_2) F_1(a_3, a_4) ]
F_2(a_5, a_6)
          \nonumber \\
& + & [ F_3(a_1, a_2) F_3(a_3, a_4) + F_4(a_1, a_2) F_2(a_3, a_4) ]
F_3(a_5, a_6)
          \nonumber \\
& + & [ F_1(a_1, a_2) F_2(a_3, a_4) + F_2(a_1, a_2) F_3(a_3, a_4) ]
F_4(a_5, a_6) \; ;
\label{ea.3}
\end{eqnarray}
\begin{eqnarray}
{1 \over 2} && \Tr [(1 - \gamma^5) \hat{a}_1 \hat{a}_2 \hat{a}_3
\hat{a}_4 \hat{a}_5 \hat{a}_6 \hat{a}_7 \hat{a}_8 ]
          \nonumber \\
= &[& F_1(a_1, a_2) F_1(a_3, a_4) + F_2(a_1, a_2) F_4(a_3, a_4) ]
          \nonumber \\
\cdot &[& F_1(a_5, a_6) F_1(a_7, a_8) + F_2(a_5, a_6) F_4(a_7, a_8) ]
          \nonumber \\[0.1cm]
+ &[& F_3(a_1, a_2) F_4(a_3, a_4) + F_4(a_1, a_2) F_1(a_3, a_4) ]
          \nonumber \\
\cdot &[& F_1(a_5, a_6) F_2(a_7, a_8) + F_2(a_5, a_6) F_3(a_7, a_8) ]
          \nonumber \\[0.1cm]
+ &[& F_3(a_1, a_2) F_3(a_3, a_4) + F_4(a_1, a_2) F_2(a_3, a_4) ]
          \nonumber \\
\cdot &[& F_3(a_5, a_6) F_3(a_7, a_8) + F_4(a_5, a_6) F_2(a_7, a_8) ]
          \nonumber \\[0.1cm]
+ &[& F_1(a_1, a_2) F_2(a_3, a_4) + F_2(a_1, a_2) F_3(a_3, a_4) ]
          \nonumber \\
\cdot &[& F_3(a_5, a_6) F_4(a_7, a_8) + F_4(a_5, a_6) F_1(a_7, a_8) ] \; ;
\label{ea.4}
\end{eqnarray}
and so forth.

\begin{equation}
\Tr [(1 + \gamma^5) \hat{a}_1 \hat{a}_2 \cdots \hat{a}_{2n} ] =
(\Tr [(1 - \gamma^5) \hat{a}_1 \hat{a}_2 \cdots \hat{a}_{2n} ])^{\ast} \; .
\label{ea.5}
\end{equation}
%


\section{Functions $F_k (a, b)$}
%
%
\begin{eqnarray}
F_1 (a, b) & = & 2 [ (a q_{-}) (b q_{+}) - (a e_{+}) (b e_{-}) ]
=  (a b) +
\begin{array}{l}
 G\pmatrix{ a & b \\ l_0 & l_1 }
\end{array}
  + i
\begin{array}{l}
 G\pmatrix{ a & b \\ l_2 & l_3 }
\end{array}
          \nonumber \\
& = & {1 \over 4} \Tr [ (1 - \gamma^5) \hat{q}_{+} \hat{q}_{-} \hat{a}
\hat{b} ]
 = - {1 \over 4} \Tr [ (1 - \gamma^5) \hat{e}_{-} \hat{e}_{+} \hat{a}
\hat{b} ] \; ,
\label{eb.1}
\end{eqnarray}
\begin{eqnarray}
F_3 (a, b) & = & 2 [ (a q_{+}) (b q_{-}) - (a e_{-}) (b e_{+}) ]
= (a b) -
\begin{array}{l}
 G\pmatrix{ a & b \\ l_0 & l_1 }
\end{array}
- i
\begin{array}{l}
 G\pmatrix{ a & b \\ l_2 & l_3 }
\end{array}
          \nonumber \\
& = & {1 \over 4} \Tr [ (1 - \gamma^5) \hat{q}_{-} \hat{q}_{+} \hat{a}
\hat{b} ]
 = - {1 \over 4} \Tr [ (1 - \gamma^5) \hat{e}_{+} \hat{e}_{-} \hat{a}
\hat{b} ] \; ,
\label{eb.2}
\end{eqnarray}
\begin{eqnarray}
F_2 (a, b) & = & 2 [ (a e_{+}) (b q_{-}) - (a q_{-}) (b e_{+}) ]
 = 2
\begin{array}{l}
 G\pmatrix{ a & b \\ e_{+} & q_{-} }
\end{array}
          \nonumber \\
& = & -
\begin{array}{l}
 G\pmatrix{ a & b \\ l_0 & l_2 }
\end{array}
  + i
\begin{array}{l}
  G\pmatrix{ a & b \\ l_1 & l_3 }
\end{array}
 +
\begin{array}{l}
 G\pmatrix{ a & b \\ l_1 & l_2 }
\end{array}
  - i
\begin{array}{l}
 G\pmatrix{ a & b \\ l_0 & l_3 }
\end{array}
          \nonumber \\
& = & {1 \over 4} \Tr [ (1 - \gamma^5) \hat{q}_{-} \hat{e}_{+} \hat{a}
\hat{b} ] \; ,
\label{eb.3}
\end{eqnarray}
\begin{eqnarray}
F_4 (a, b) & = & 2 [ (a e_{-}) (b q_{+}) - (a q_{+}) (b e_{-}) ] = 2
\begin{array}{l}
 G\pmatrix{ a & b \\ e_{-} & q_{+} }
\end{array}
          \nonumber \\
& = & -
\begin{array}{l}
 G\pmatrix{ a & b \\ l_0 & l_2 }
\end{array}
+ i
\begin{array}{l}
G\pmatrix{ a & b \\ l_1 & l_3 }
\end{array}
-
\begin{array}{l}
 G\pmatrix{ a & b \\ l_1 & l_2 }
\end{array}
+ i
\begin{array}{l}
 G\pmatrix{ a & b \\ l_0 & l_3 }
\end{array}
          \nonumber \\
& = & {1 \over 4} \Tr [ (1 - \gamma^5) \hat{q}_{+} \hat{e}_{-} \hat{a}
\hat{b} ] \; .
\label{eb.4}
\end{eqnarray}
\begin{equation}
F_5 (a, b) = F_1^{\ast} (a, b)  , \; F_6 (a, b) = F_2^{\ast} (a, b)  , \;
F_7 (a, b) = F_3^{\ast} (a, b)  , \; F_8 (a, b) = F_4^{\ast} (a, b)  .
\label{eb.5}
\end{equation}
%
%

\end{appendix}




\end {document}